\documentclass[aps,pre,twocolumn,superscriptaddress,groupedaddress]{revtex4-1}  

\usepackage{cmap}  
\usepackage[T1]{fontenc}
\usepackage[latin1]{inputenc}
\usepackage[english]{babel}

\usepackage{amsmath, amssymb, amsthm}  
\usepackage{bm, bbm}

\usepackage[]{graphicx}

\usepackage[separate-uncertainty=true,multi-part-units=single]{siunitx}
\usepackage{enumerate} 
\usepackage{dcolumn}   
\usepackage{lineno}

\usepackage[colorlinks = true, linkcolor = blue, urlcolor  = blue, citecolor = blue,       anchorcolor = blue]{hyperref} 

\usepackage{comment}
\usepackage{slashed}    
\usepackage{mathtools}  

\usepackage{tikz}
\usepackage{pgfplots} 
\usepackage{pgfgantt}
\pgfplotsset{compat=newest} 
\pgfplotsset{plot coordinates/math parser=false}
\usetikzlibrary{external}
\newlength\fwidth
\usetikzlibrary{plotmarks}

\DeclareMathOperator{\diag}{diag}

\let\Re\relax\DeclareMathOperator{\Re}{Re}

\newcommand{\iu}{\mathfrak{i}}
\newcommand{\ex}{e}
\newcommand{\du}{d}

\newcommand{\mat}[1]{\bm #1}
\renewcommand{\vec}[1]{\bm #1}

\newcommand{\uv}[1]{\bm{\mathrm{#1}}}           
\newcommand{\mean}[1]{\langle #1 \rangle}

\newcommand{\abs}[1]{\lvert #1 \lvert}
\newcommand{\fra}[1]{\frac{1}{#1}}                          
\newcommand{\tfra}[1]{\tfrac{1}{#1}}

\addto\captionsenglish{}
\addto\captionsenglish{}

\newcommand{\Lval}{\SI{0.80}{\milli\meter}}
\newcommand{\Lvalue}{\SI{0.80\pm 0.24}{\milli\meter}}

\newcommand{\DAval}{\SI{0.48}{\centi\meter\squared\per\second}}
\newcommand{\DAvalue}{\SI{0.48\pm 0.12}{\centi\meter\squared\per\second}}

\newcommand{\Rval}{\SI{0.35}{\milli\meter}}
\newcommand{\Rvalue}{\SI{0.35\pm 0.04}{\milli\meter}}

\newcommand{\DBval}{\SI{0.86}{\centi\meter\squared\per\second}}

\begin{document}

\title{Analytic determination of lung microgeometry with gas diffusion magnetic resonance}
\author{Niels Buhl}
\email{nbuhl@phys.au.dk}
\affiliation{School of Physics and Astronomy, University of Nottingham, Nottingham NG7 2RD, United Kingdom}


\begin{abstract}
Through inhalation of, e.g., hyperpolarized $^3$He, it is possible to acquire gas diffusion magnetic resonance measurements that depend on the local geometry in the vast network of microscopic airways that form the respiratory zone of the human lung.      
Here, we demonstrate that this can be used to determine the dimensions (length and radius) of these airways noninvasively.
Specifically, the above technique allows measurement of the weighted time-dependent diffusion coefficient (also called the apparent diffusion coefficient), which we here derive in analytic form using symmetries in the airway network. 
Agreement with experiment is found for the full span of published hyperpolarized $^3$He diffusion magnetic resonance measurements (diffusion times from milliseconds to seconds) and published invasive airway dimension measurements.       
\end{abstract}

\date{\today}
\maketitle 

\section{Introduction}
Ever since the microgeometry dependence of gas diffusion magnetic resonance (MR) measurements on the lung was first demonstrated experimentally  
\cite{Chen2000,Saam2000,Salerno2002}, there has been great interest in utilizing this dependence in clinical and research applications, such as detection and investigation of  emphysema          
(see, e.g., Refs.~\cite{Yablonskiy2002,Yablonskiy2009,Wang2008,Trampel2006,Kirby2012,Evans2007,Ouriadov2018,Fain2006,Altes2006,Stewart2018,Chan2017-3He,Morbach2005,van-Beek2009,Narayanan2012,Wang2008method,Wang2006,Gonen2016,Yablonskiy2017,Westcott2019}).  
However, due to the lack of a general analytic relation relating the microgeometry and the measured quantity, it has so far not been clear what specific information the dependence allows one to extract. Here, we provide this missing piece, which shows that the dependence allows extraction of the dimensions of the respiratory-zone airways.

\begin{figure}[htp]
\centering
\setlength\fwidth{0.7\textwidth} %
\includegraphics{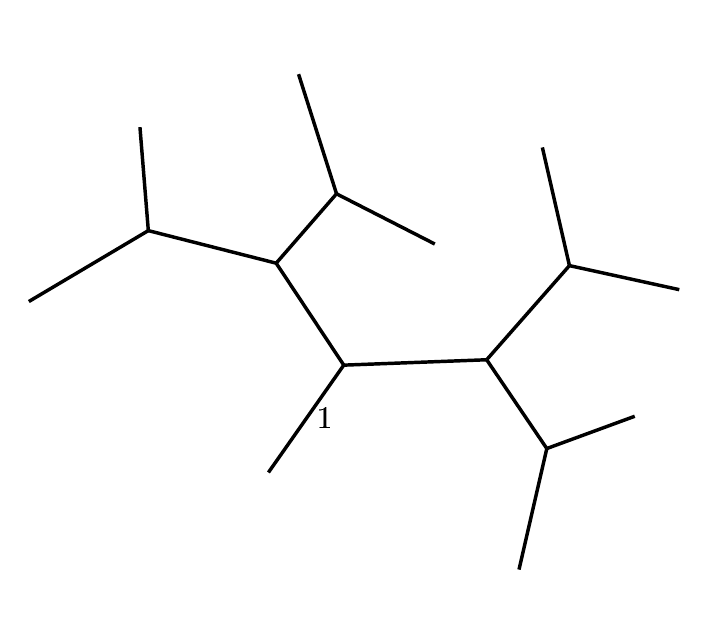}
\caption{\label{fig:acinus}Schematic illustration of the airway branching pattern in a pulmonary acinus. 
The line segments represent the center-line segments of the airways that make up the first (labeled $1$ in the figure), second, third, and fourth of the nine generations of airways in the pulmonary acinus.    
}
\end{figure}

The human lung has about $24$ generations of airways arranged such that each airway in a given nonlast generation (counting the trachea as the first generation) branches into two airways in the subsequent generation \cite{Weibel1997}.     
The airways that make up the last about nine generations all have a central tubular region with radial openings into cavities formed by a surrounding sleeve of alveoli  \cite{Weibel1997}.    
These airways, which together form the respiratory zone and together account for about $95\%$ of the lung's airspace, have roughly the same length \cite{Haefeli-Bleuer1988,Litzlbauer2010}, roughly the same radius measured from the centerline to and including the alveolar sleeve \cite{Haefeli-Bleuer1988,Litzlbauer2010}, and roughly the same sibling-to-sibling angle of about \ang{120} \cite{Litzlbauer2010}. 
A respiratory-zone airway that immediately succeeds a non-respiratory-zone airway forms, together with all its descendants, a so-called pulmonary acinus (see Fig.~\ref{fig:acinus}).

The technique of gas diffusion MR with, e.g., inhaled hyperpolarized $^3$He \cite{Gentile2017} allows regional and global measurement of the weighted time-dependent diffusion coefficient (also called the apparent diffusion coefficient) with a diffusion time in the millisecond to second range (for which the corresponding characteristic free-diffusion length is on the order of $1$ to $10$ airway lengths in the respiratory zone) \cite{Wang2008,Wang2008method,Wang2006}.  
Deriving the above quantity in analytic form, is consequently not possible by considering the airways as non-interconnected infinite-length objects, as in the fixed-diffusion-time framework of Refs.~\cite{Yablonskiy2002,Yablonskiy2009} see also the review Ref.~\cite{Yablonskiy2017}, but requires taking into account the airways' finite length and interconnections.

In this work, we derive in analytic form the lung's weighted time-dependent diffusion coefficient based on considering the pulmonary acinus as the representative unit.  
First, we show that this  quantity splits into two parts, one that can be handled by considering the diffusive motion of the gas molecules' projections on the geometric graph formed by the airways' center-line segments and one that can be handled by considering the transverse part of the gas molecules' diffusive motion in each airway.  
By using symmetries in the airway network together with this, we get the above quantity expressed analytically in terms of the airway length and the airway radius.
Agreement with experiment is found for the full span of published hyperpolarized $^3$He diffusion MR measurements and published invasive airway dimension measurements.   
\vspace{0.1cm}

\section{Theory and Results}
Let $S$ be the magnitude of the (gradient echo, pulsed gradient spin echo, or pulsed gradient stimulated echo) diffusion-sensitized MR signal  from the volume of interest at the time of the echo.   
Let $\Delta$ be the diffusion time (i.e., the inter-pulse time), $\delta$ be the pulse duration, and $T$ be the time of the echo. 
Let $G(t)\uv g$ be the effective field gradient for the diffusion sensitization such that $\uv g$ is a unit vector. We then have $\int_0^T G(t)dt=0$ and  
\begin{align*}
S=S_0\mean{\exp(-\iu\varphi)},
\qquad \varphi=\int_{0}^{T}\gamma G(t)\uv g\cdot\vec r(t)\du t,
\end{align*}
where $S_0$ is the magnitude of the corresponding  non-diffusion-sensitized MR signal from the volume of interest at time $T$, $\gamma$ is the gyromagnetic ratio, $\vec r(t)$ is the position vector of the $i$th gas molecule, and $\mean{\cdot}$ denotes averaging over all gas molecules in the volume of interest.  
By expanding in cumulants we get the equation       $\ln(S/S_0)=-\tfra2\mean{\varphi^2}+\tfra{24}(\mean{\varphi^4}-3\mean{\varphi^2}^2)+\dotsb$ 
which, by evaluating   $\tfra2\mean{\varphi^2}$,  gives   
\begin{gather*}
\ln(S/S_0)=-\mathfrak D_{\uv g}b+ O(b^2),
\end{gather*}
where $b=-\gamma^2\int_0^{T}\int_0^{t}  G(t)G(s)(t-s)\du s\du t$ 
(this expression for $b$ is equal to the well-known  expression     
$\gamma^2\int_0^T(\int_0^t  G(s)ds)^2dt$ as can be seen by integration by parts) and  $\mathfrak D_{\uv g}$ is equal to the expression we get by substituting $\mean{(\uv g\cdot(\vec r(t)-\vec r(0)))^2}/2t$ for $\mathcal D(t)$ in \eqref{eq:WDCdef}. Thus, $\mathfrak D_{\uv g}$ is  equal to the initial slope of, the $\ln(S_0/S)$ versus $b$ curve obtained by varying the strength of $G(t)$ while keeping the timing parameters of $G(t)$ unchanged.      
By further varying the direction of the diffusion sensitization, i.e., $\uv g$, over three orthonormal directions given by $\uv g_1$, $\uv g_2$, and $\uv g_3$      we get   
 $\mathfrak D=(\mathfrak D_{\uv g_1}+\mathfrak D_{\uv g_2}+\mathfrak D_{\uv g_3})/3$, where $\mathfrak D$ is the weighted time-dependent diffusion coefficient    
\begin{gather}\label{eq:WDCdef}
\mathfrak D=\frac{\int_0^{T}\int_0^{t} G(t) G(s)(t-s)\mathcal D(t-s)\du s\du t}
{\int_0^{T}\int_0^{t} G(t)G(s)(t-s)\du s\du t},
\end{gather} 
and $\mathcal D(t)$ is the time-dependent diffusion coefficient  
\begin{gather}\label{eq:D(t)} 
\mathcal D(t)=\frac{\mean{(\vec r(t)-\vec r(0))^2}}{6t}.
\end{gather}

To derive $\mathfrak D$ in analytic form, we consider, as a representative portion of the volume of interest, a pulmonary acinus with $\nu$, $\nu=9$, complete generations and thus a total of $2^\nu-1$ airways. 
Let $\theta$, $\theta=\pi/3$, be the airway branching angle and $2\theta$ be the airway sibling-to-sibling angle.
Let $L$ be the airway length and $R$ be the airway radius measured from the centerline to and including the alveolar sleeve.  
Let the airways be numbered such that airway $1$ is the airway from which all the other airways descend and if airway $i$ has children, then airway $i$'s two children are airway $2i$ and airway $2i+1$. 
Let $\vec o_i$ be the position vector and $\uv x_i$, $\uv y_i$, and $\uv z_i$ be orthogonal unit vectors such that $\vec o_i$ and $\vec o_i+L\uv x_i$ are the positions of the parent-end end point and the non-parent-end end point, respectively, of airway $i$'s center-line segment. 
Let $l_t$, $x_t$, $y_t$, and $z_t$ be such that $\vec r(t)=\vec o_{l_t}+x_{t}\uv x_{l_t}+y_t\uv y_{l_t}+z_t\uv z_{l_t}$ (the $t$-dependence is here denoted using a subscript to ease the notation).   
By inserting this in \eqref{eq:D(t)} and expanding we get terms such as  $\tfrac{1}{6t}\mean{-2y_t\uv y_{l_t}\cdot x_0\uv x_{l_0}}$.  
This term is equal to  $\tfra{6t}\sum_{i,i'}w_{i'i}\uv y_{i'}\cdot\uv x_i$, where $w_{i'i}\uv y_{i'}\cdot\uv x_i$ is that part of the average that concerns the gas molecules that are in airway $i$ initially and in airway $i'$ at time $t$. 
Since a gas molecule can only diffuse from one airway to another by passing through the central tubular regions of these, we have that $y_t$ is as likely to be positive as negative if $l_t\ne l_0$ and hence that $w_{i'i}=0$ if $i'\ne i$.
Consequently we have  $\tfra{6t}\mean{-2y_t\uv y_{l_t}\cdot x_0\uv x_{l_0}}=0$. 
By applying the same reasoning to the other terms in the expansion we get     
\begin{align*}
\mathcal D(t)=&\tfra{6t}\mean{(\vec o_{l_t}+x_t\uv x_{l_t}-\vec o_{l_0}-x_0\uv x_{l_0})^2}\nonumber\\
&\phantom{..}+\tfra{6t}\mean{(y_t\uv y_{l_t}+z_t\uv z_{l_t}-y_0\uv y_{l_0}-z_0\uv z_{l_0})^2}.
\end{align*}
The first part of $\mathcal D(t)$ can be expressed analytically by using Ref.~\cite{Buhl2018} to obtain in analytic form the continuum-limit propagator for the diffusive motion of the gas molecules' projections on the geometric graph formed by the airways' center-line segments. 
By further using the transverse part of the cylindrical-coordinate  continuum-limit propagator for particle diffusion in a nonabsorbing hollow cylinder with radius $R$ (see, e.g., Refs.~\cite{Neuman1974,Yablonskiy2002}) to express the second part of $\mathcal D(t)$ analytically we get after inserting the sum of the two expressions in \eqref{eq:WDCdef}      
\begin{gather*}
\mathfrak D=
\sum_{n=1}^\infty
a_n L^2f(-\alpha_n^2 D_a/L^2)+\sum_{n=1}^\infty b_nR^2 f(-\beta_{1n}^2D_b/R^2),
\end{gather*}
where $\alpha_n$ and $a_n$ are given in Appendix~\hyperref[app:A]{A},  $\beta_{1n}$ is the $n$th root of the derivative of the $1$th-order Bessel function of the first kind, $b_n=4/(3\beta_{1n}^2 -3\beta_{1n}^4)$, $D_a$ is the diffusion coefficient for the diffusive motion of the gas molecules' projections on the geometric graph, $D_b$ is the (independently measurable) diffusion coefficient for the diffusive motion of the gas molecules in the absence of confinement, and $f(\kappa)$ is equal to the expression we get by substituting $\exp(\kappa (t-s))$ for $(t-s)\mathcal D(t-s)$ in \eqref{eq:WDCdef}. 
For $G(t)$ consisting of two rectangular pulses (i.e., being equal to      $g\Pi({0,\delta},t)-g\Pi({\Delta,\Delta+\delta},t)$ up to translation and  signs, where  
$\Pi({a,b},t)=H(t-a)-H(t-b)$ is the boxcar function, $H(t)$ is the Heaviside step function, and $g$ is the strength of $G(t)$) we have      
 \begin{align*}
f(\kappa)=\frac{2\kappa\delta+2-2\ex^{\kappa\Delta}-2\ex^{\kappa\delta}+\ex^{\kappa(\Delta-\delta)}+\ex^{\kappa(\Delta+\delta)}}{\kappa^2\delta^2(\Delta-\delta/3)}.
\end{align*}

\begin{figure}[htp]
\centering
\setlength\fwidth{0.4\textwidth}
\includegraphics{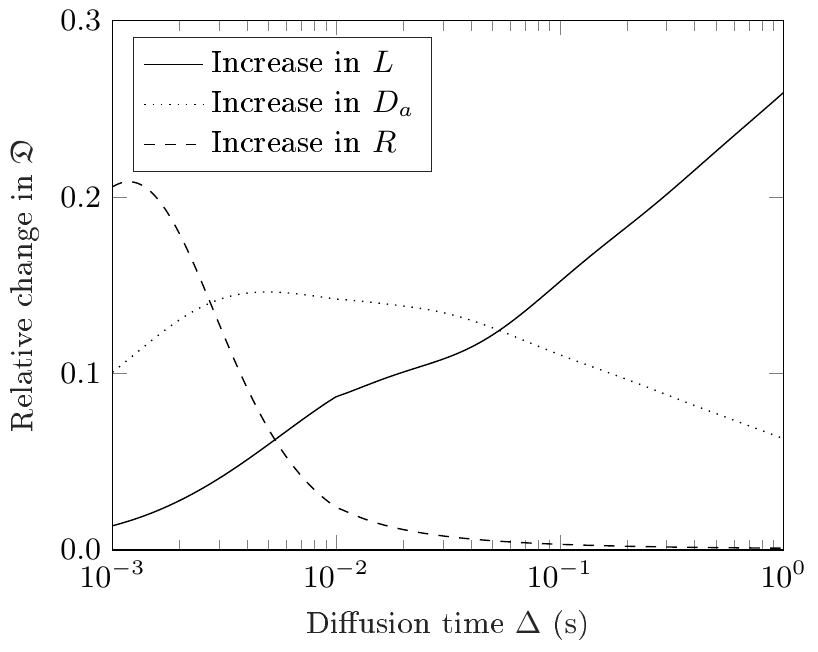}
\caption{\label{fig:sensitivity}Illustration of how the sensitivity of $\mathfrak D$ to changes in $L$, $D_a$, and $R$ varies with $\Delta$. The solid, dotted, and dashed curve shows the relative change in $\mathfrak D$ caused by a $20\%$ increase in $L$, $D_a$, and $R$, respectively, computed with the following parameter values as reference: $L=\Lval$, $D_a=\DAval$, $R=\Rval$, $D_b=\DBval$,  $G(t)$ consisting of two rectangular pulses, and $\delta$ equal to $\Delta$ if $\Delta\le \SI{10}{\milli\second}$ and equal to \SI{10}{\milli\second} otherwise.
}  
\end{figure}

The sensitivity of $\mathfrak D$ to changes in $L$ and $R$ will because of the difference between      $(\alpha_1^2,\alpha_2^2,\dotsc)$ and $(\beta_{11}^2,\beta_{12}^2,\dotsc)$ vary differently with $\Delta$ (the former and latter sequence arranged in ascending order are numerically        $(0.0020,0.0042,\dotsc)$ and $(3.3900, 28.4243,\dotsc)$, respectively). 
For values of $\Delta$ such that $D_a\Delta/L^2\sim 1$ and $D_b\Delta/R^2\sim 1$ we have, for $G(t)$ consisting of two rectangular pulses and $\delta=\Delta$, that $\mathfrak D$ is approximately equal to                  
$\frac{D_a}{3}\big(1-\tfrac{257}{511 L}\tfrac{32(2\sqrt2-1)}{35\sqrt{\pi}}(D_a\Delta)^{1/2}\big) 
+\tfrac{7}{48}\frac{R^4}{D_b\Delta^2}-\tfrac{33}{512}\frac{R^6}{D_b^2\Delta^3}$ 
(see Appendix~\hyperref[app:B]{B} and Fig.~\ref{fig:shorttime}) 
and for values of $\Delta$ such that $D_a\Delta/L^2\gg 1$ and $D_b\Delta/R^2\gg 1$ we have that the second part of $\mathfrak D$ is negligible compared with the first part. Thus, the sensitivity of $\mathfrak D$ to changes in $L$ and $R$ goes from moderate to high and high to low, respectively, if $\Delta$ is increased from the first to the second of these two regimes, see Fig.~\ref{fig:sensitivity}. 
Note, that the above approximation is linear with respect to four expressions  that together determine $L$, $D_a$, $R$, and $D_b$. 
 
\begin{figure}[htp]
\centering
\setlength\fwidth{0.4\textwidth}
\includegraphics{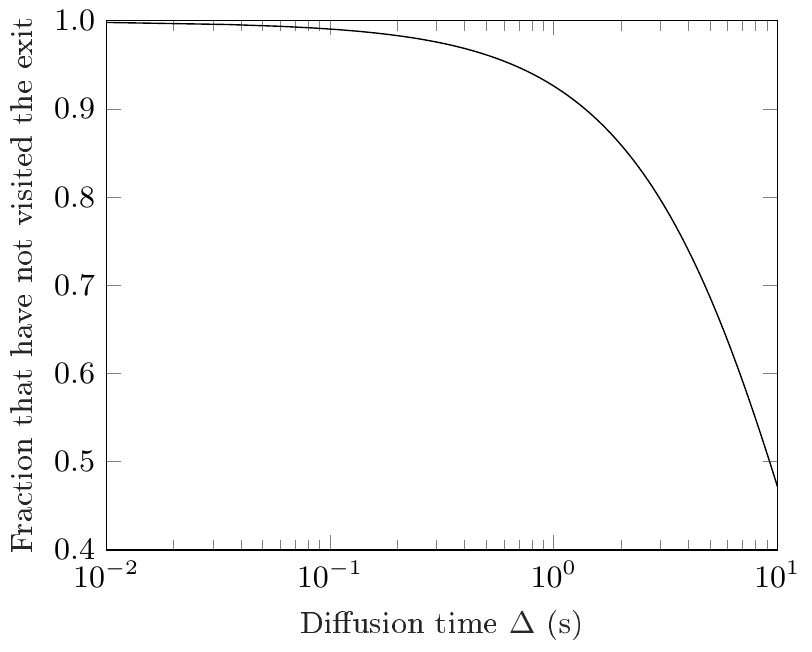}
\caption{\label{fig:fraction}Fraction of the gas molecules in the pulmonary acinus that have not visited its exit as a function of $\Delta$. Specifically, the curve shows $\fra{EL}\sum_{i,i'}\int_0^L\!\!\int_0^L\bar K(x',x,\Delta)_{i'i}\du x\du x'$ as a function of $\Delta$ computed with $L=\Lval$ and $D_a=\DAval$, where the integrand, which can be obtained using Ref.~\cite{Buhl2018}, is the continuum-limit propagator for the diffusive motion of the gas molecules' projections on the geometric graph formed by the airways' center-line segments subject to the condition that the vertex located at $\vec o_1$ is absorbing and all other vertices are nonabsorbing.   
}
\end{figure}

\begin{figure}[htp]
\centering
\setlength\fwidth{0.4\textwidth}
\includegraphics{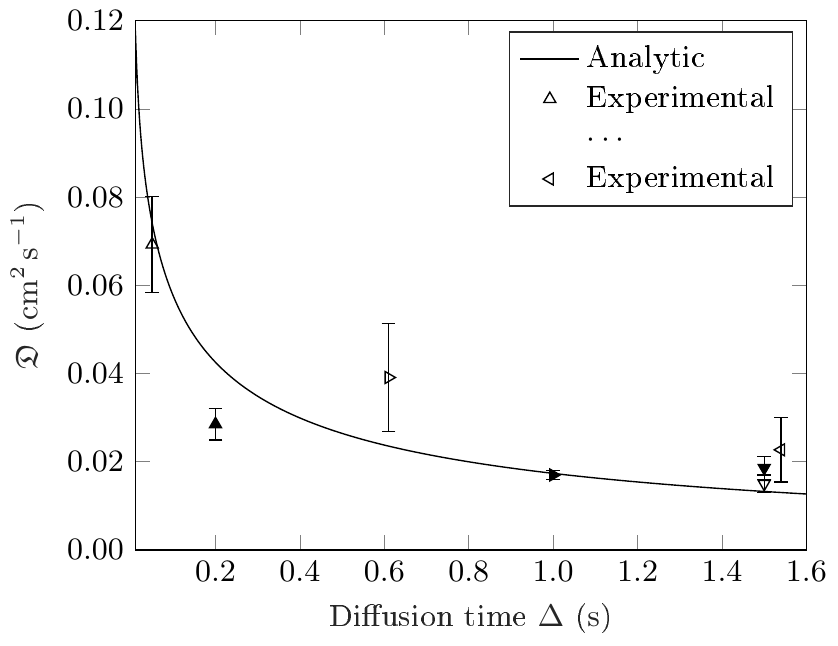}
\caption{\label{fig:mediumlongtime}$\mathfrak D$ as a function of $\Delta$ computed and experimentally estimated ($\Delta$ varied in the range \SIrange{0.01}{1.6}{\second}). The solid curve shows $\mathfrak D$ as a function of $\Delta$ computed with $L=\Lval$, $D_a=\DAval$, $R=\Rval$, $D_b=\DBval$, $G(t)$ consisting of two rectangular pulses, and  $\delta$ equal to $\Delta$ if $\Delta\le \SI{10} {\milli\second}$ and equal to \SI{1}{\milli\second} otherwise. The symbols show $\mathfrak D$ as a function of $\Delta$ experimentally estimated based on the hyperpolarized $^3$He diffusion MR measurements on healthy adults reported in Refs.\ (each symbol and error bar represents the mean and standard deviation of measurements on $n$ healthy adults)   
\cite{Wang2008method} (\protect\includegraphics{mark1}, $n=4$), 
\cite{Wang2008method} (\protect\includegraphics{mark2}, $n=2$), 
\cite{Wang2006} (\protect\includegraphics{mark3}, $n=10$),
\cite{Gonen2016} (\protect\includegraphics{mark4}, $n=17$), 
\cite{Wang2008} (\protect\includegraphics{mark5}, $n=14$),   
\cite{Wang2008method} (\protect\includegraphics{mark6}, $n=3$), and 
\cite{Wang2006} (\protect\includegraphics{mark7}, $n=10$). 
}
\end{figure}

To see that there is agreement with experiment, we consider the full span of published hyperpolarized $^3$He MR measurements on healthy adults (gas diffusion MR measurements on another population or with another gas, such as hyperpolarized $^{129}$Xe \cite{Norquay2018}, that form a multi-diffusion-time dataset are not yet available).
These measurements are made with either a gradient echo sequence with $\Delta$ in the range \SIrange{1}{5.2}{\milli\second} and $\delta\approx\Delta$ or a pulsed gradient stimulated echo sequence with $\Delta$ in the range \SIrange{0.05}{1.54}{\second} and $\delta$ roughly on the order of \SI{1}{\milli\second}. 
For the $^3$He concentrations used in these measurements, $D_b$ has a measured value of about $\DBval$ (see Ref.~\cite{Acosta2006} for the functional form of the concentration dependence).  
The symbols in Figs.~\ref{fig:shorttime} and \ref{fig:mediumlongtime} show $\mathfrak D$ as a function of $\Delta$ experimentally estimated based on these measurements.     
To assess the representativeness of considering only gas diffusion restricted to within the pulmonary acinus for the diffusion times used in these measurements, we compute in Fig.~\ref{fig:fraction} the fraction of the gas molecules in the pulmonary acinus that have not visited its exit, as a function of $\Delta$, from which it can be seen that this fraction for all the diffusion times used in these measurements is greater than about $0.9$.   
Least squares fitting the analytic form of $\mathfrak D$, with $L$, $D_a$, and $R$ as adjustable parameters, $D_b=\DBval$, $G(t)$ consisting of two rectangular pulses, and  $\delta$ equal to $\Delta$ if $\Delta\le \SI{10}{\milli\second}$ and equal to \SI{1}{\milli\second} otherwise, to the experimental estimate represented by the symbols in Figs.~\ref{fig:shorttime} and \ref{fig:mediumlongtime} gives $L=\Lvalue$  $D_a=\DAvalue$, and $R=\Rvalue$. 
The solid curve in Figs.~\ref{fig:shorttime} and \ref{fig:mediumlongtime} shows $\mathfrak D$ as a function of $\Delta$ computed with these parameter values.
The above result for $L$ and $R$ is consistent with the published invasive findings. 
Specifically, Ref.~\cite{Haefeli-Bleuer1988} found $L$ to be \SI{0.871}{\milli\meter} (the mean of the means in columns 2 and 4 of Table~2) and $R$ to be \SI{0.339}{\milli\meter} (the mean of the means in columns 10 and 12 of Table~2) based on microscope examination of corrosion casts of two fixed normal adult human lungs and Ref.~\cite{Litzlbauer2010} found $L$ to be \SI{0.638}{\milli\meter} and $R$ to be \SI{0.375}{\milli\meter} based on synchrotron-based micro-computed tomography of osmium tetroxide stained samples of a fixed normal adult human lung.

\begin{figure}[htp]
\centering
\setlength\fwidth{0.4\textwidth}
\includegraphics{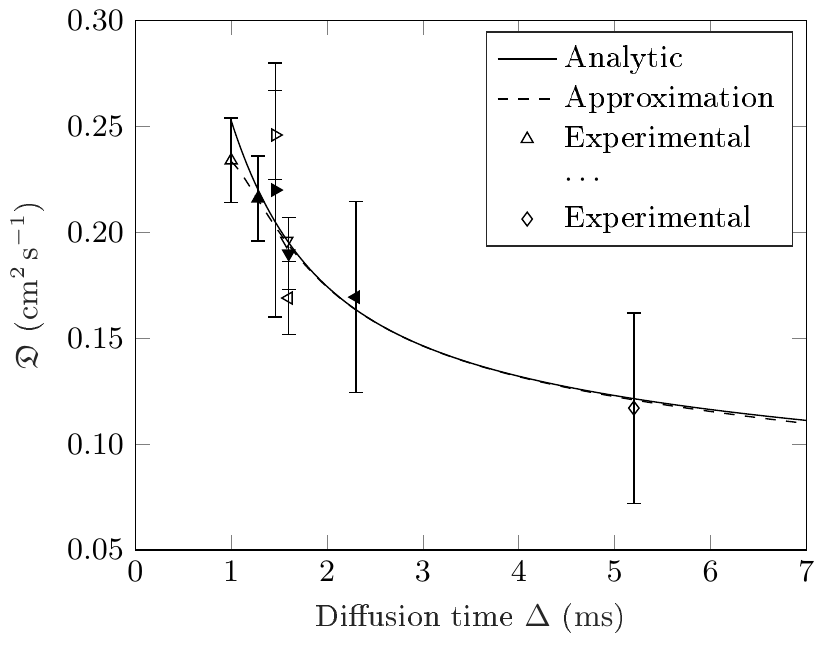}
\caption{\label{fig:shorttime}$\mathfrak D$ as a function of $\Delta$ computed and experimentally estimated ($\Delta$ varied in the range \SIrange{1}{7}{\milli\second}). The solid curve shows ${\mathfrak D}$ as a function of $\Delta$ computed with $L=\Lval$, $D_a=\DAval$, $R=\Rval$, $D_b=\DBval$, $G(t)$ consisting of two rectangular pulses, and  $\delta$ equal to $\Delta$ if $\Delta\le \SI{10}{\milli\second}$  and equal to \SI{1}{\milli\second} otherwise. 
The dashed curve shows the $D_a\Delta/L^2\sim 1$ and $D_b\Delta/R^2\sim 1$ approximation for $\mathfrak D$ as a function of $\Delta$ computed with the same parameter values.
The symbols show $\mathfrak D$ as a function of $\Delta$ experimentally estimated based on the hyperpolarized $^3$He diffusion MR measurements on healthy adults reported in Refs.\ (each symbol and error bar represents the mean and standard deviation of measurements on $n$ healthy adults)     
\cite{Wang2008} (\protect\includegraphics{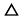}, $n=14$),
\cite{Trampel2006} (\protect\includegraphics{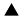}, $n=4$), 
\cite{Kirby2012} (\protect\includegraphics{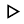}, $n=8$, see also Ref.~\cite{Evans2007}), 
\cite{Ouriadov2018} (\protect\includegraphics{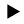}, $n=15$, see also Ref.~\cite{Fain2006}), 
\cite{Altes2006} (\protect\includegraphics{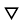}, $n=8$), 
\cite{Stewart2018} (\protect\includegraphics{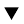}, $n=11$),
\cite{Chan2017-3He} (\protect\includegraphics{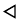}, $n=5$), 
\cite{Morbach2005} (\protect\includegraphics{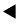}, $n=6$, see also Ref.~\cite{van-Beek2009}), and 
\cite{Narayanan2012} (\protect\includegraphics{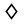}, $n=5$, the authors of \cite{Narayanan2012} kindly made the raw gas diffusion MR data available).    
}
\end{figure}

\section{Discussion and Conclusion}
In this work, we showed how the lung's weighted time-dependent diffusion coefficient depends on the lung microgeometry and how the sensitivity characteristics of the dependence vary with the diffusion time. 
Hereby, we showed how the dimensions (length and radius) of the respiratory-zone airways can be extracted, from knowledge of the value of the weighted time-dependent diffusion coefficient as a function of the diffusion time.  
The agreement with experiment of this was demonstrated for the full span of published hyperpolarized $^3$He diffusion MR measurements and published invasive airway dimension measurements.    
As directions for future use we note the importance of ensuring that it is indeed the initial slope of the $\ln(S_0/S)$ versus $b$ curve that is evaluated when measuring the weighted time-dependent diffusion coefficient.   
Also, the diffusion times used should not be greater than what can ensure the representativeness of considering only gas diffusion restricted to within the pulmonary acinus (as assessed in terms of the fraction of the gas molecules in the pulmonary acinus that have not visited its exit, see Fig.~\ref{fig:fraction} and its caption). 
Thus, this work opens the door to gas diffusion-MR-based examination of the human lung, where extraction of the relevant dimensions of the underlying lung microgeometry is \hbox{possible}.

\section*{Acknowledgment}
This work was partially funded by the FP7 EU grant-funded project AirPROM.

\section*{Appendix A}\label{app:A}
\setcounter{equation}{0} 
\makeatletter
\renewcommand{\theequation}{A\arabic{equation}} %
In this Appendix, we show in detail how to obtain $\mathfrak D$ in analytic form.  
First, we use Ref.~\cite{Buhl2018} to obtain in analytic form the continuum-limit propagator for the diffusive motion of the gas molecules' projections on the geometric graph formed by the airways' center-line segments (subject to the condition that all vertices are nonabsorbing). The resulting analytic expression is as follows:     
Let the edges and vertices of  the geometric graph be numbered such that edge $i$ is the center-line segment of airway $i$, the vertex located at $\vec o_i$ is vertex $\phi_i$, and the vertex located at $\vec o_i+L\uv x_i$ is vertex $\psi_i$, where $\phi_i=\lfloor i/2\rfloor+1$ and $\psi_i=i+1$. 
Let $\mat{\Phi}$ and $\mat{\Psi}$ be the two $E\times V$ matrices defined by  
\begin{align*}
\mat{\Phi}_{ij}=\delta_{\lfloor i/2\rfloor+1, j},\qquad\mat{\Psi}_{ij}=\delta_{i+1,j},
\end{align*}
where $E=2^\nu-1$ is the number of edges, $V=2^\nu$ is the number of vertices,  $\delta_{ij}$ is the Kronecker delta, and   $\lfloor\cdot\rfloor$ is the floor function.  
Let
\begin{align*}
\mat D&=\mat\Phi^\top\mat{\Phi}+\mat{\Psi}^\top\mat{\Psi},\\
\mat A&=\mat\Phi^\top\mat{\Psi}+\mat{\Psi}^\top\mat{\Phi}.
\end{align*}
Let $\mat R=\mat D^{-1/2}$ (note that $\mat D$ is diagonal) and $\mat W=\mat{ R}\mat A\mat R$. 
Let $\mat V$ and $\mat X$ be real matrices of the same size as $\mat W$ such that $\mat W\mat V=\mat V\mat X$, $\mat V^\top\mat V=\mat I$, and $\mat X$ is diagonal with the diagonal entries arranged in ascending order (these matrices can be found (using standard mathematical software) since $\mat W$ is real symmetric and therefore real orthogonally diagonalizable).   
Let $\hat{\mat V}_-$   be the first column in $\mat R\mat V$ and $\hat{\mat V}_+$ be the last column in $\mat R\mat V$. Let $\hat{\mat V}$ be the matrix we get by deleting the first and the last column from $\mat R\mat V$. Let $\hat{\mat X}$ be the matrix we get by deleting the first and the last column as well as the first and the last row from $\mat X$. Let  $\mat\Lambda$ be the matrix of the same size as $\hat{\mat X}$ defined by    $\mat\Lambda_{ij}=\arccos(\hat{\mat X}_{ii})\delta_{ij}$  
(the entries of $\hat{\mat X}$ are strictly between $-1$ and $1$).  
Let    
\begin{align*}
\mat\Theta&=\diag({\mat\Lambda}, \pi, 2\pi\mat I-\mat\Lambda, 2\pi),\\
\mat P&=\left[\begin{array}{cccccc}(\mat\Phi\mat Y-\mat\Psi\mat Y\mat Z) & \mat\Phi\hat{\mat V}_- & (\mat\Phi\mat Y-\mat\Psi\mat Y\mat Z^\ast) & \mat\Phi\hat{\mat V}_{+} \end{array}\right],
\\\mat Q&=\left[\begin{array}{cccccc}(\mat\Psi\mat Y-\mat\Phi\mat Y\mat Z) & \mat\Psi\hat{\mat V}_- & (\mat\Psi\mat Y-\mat\Phi\mat Y\mat Z^\ast) & \mat\Psi\hat{\mat V}_{+} \end{array}\right],
\end{align*}
where $\mat Y=\iu\hat{\mat V}(2\mat I-2\hat{\mat X}^2)^{-1/2}$  
and $\mat Z=\hat{\mat X}+\iu(\mat I-\hat{\mat X}^2)^{1/2}$. Throughout, $\iu$ is the imaginary unit, $^\ast$ denotes complex conjugation, and $\mat I$ is the identity matrix whose size is determined by the context.
The continuum-limit propagator is then
\begin{align*}
\bar K(x',x,t)_{i'i}=\fra{EL}+\sum_{n=1}^\infty\ex^{-\alpha_n^2D_a t/L^2}u_{{i'} n}(x')u_{i n}(x), 
\end{align*}
where 
\begin{gather*}
\alpha_{n}=\lfloor(n-1)/(2E)\rfloor 2\pi +\mat\Theta_{r_n r_n},\\ 
u_{in}(x)=2L^{-1/2}\Re(\mat Q_{i r_n}\ex^{-\iu\alpha_nx/L}),
\end{gather*}
in which    $r_n=n-\lfloor(n-1)/(2E)\rfloor 2E$ and $\Re(z)$ is the real part of $z$.  
With the above we can express the first part of $\mathcal D(t)$ as  
\begin{align*}
&\fra{6t}\mean{{(\vec o_{l_t}+x_t\uv x_{l_t}-\vec o_{l_0}-x_0\uv x_{l_0})^2}}\\
&\phantom{.}=\frac{1}{6t}
\sum_{i,i'}\int_0^L\!\!\int_0^L\fra{EL}\bar{K}(x',x,t)_{i'i}\\
&\phantom{=.}\times(L\sum_{k}\mat O_{ki'}\uv x_{k}+x'\uv x_{i'}-L\sum_{k}\mat O_{ki}\uv x_k-x\uv x_{i})^2\du x\du x',
\end{align*}
where $\mat O$ is the $E\times E$ matrix such that $\mat O_{ki}$ is equal to $1$ if airway $k$ is an ancestor of airway $i$ and equal to $0$ otherwise and thus $\vec o_i=\vec o_1+L\sum_{k}\mat O_{ki}\uv x_{k}$.   
The above double sum can be divided into subsums of the form $\sum_{(i,i')\in A(m,n)}$ or $\sum_{(i,i')\in B(m,n,u)}$, where $(i,i')\in A(m,n)$ if and only if airway $i$ is identical, an ancestor, or a descendant of airway $i'$, the former belongs to generation $m$, and the latter belongs to generation $n$ and $(i,i')\in B(m,n,u)$ if and only if airway $i$ is not identical, not an ancestor, and not a descendant of airway $i'$, the former belongs to generation $m$, the latter belongs to generation $n$, and the lowest common ancestor to the former and the latter belongs to generation $u$. 
By using this and expanding the factor $(\dotsb)^2$ we get an expression that is equal to the above and consist of terms such as $-\tfra{6tEL}\sum_{(i,i')\in A(m,n)}\int_0^L\!\!\int_0^L \bar{K}(x',x,t)_{i'i}2x'x\du x\du x'\uv x_{i'}\cdot\uv x_i$.   
Because of the symmetries in the airway network, all the double integrals in this sum have the same value, and the sum can thus be computed by using that 
\begin{align}\label{eq:sumxixj}
&\sum_{{(i,i')\in A(m,n)}}\uv x_{i'}\cdot\uv x_{i}
=\sum_{{(i,i')\in A(m,n)}}\mat C_{i'i},
\end{align}
where $\mat C$ is the $E\times E$ matrix ${\mat C_{ij}=\prod_{k=1}^{m-1}\uv x_{\ell_k}\cdot\uv x_{\ell_{k+1}}}$ in which $\ell_1,\dotsc,\ell_m$ 
are such that $(\ell_1,\dotsc,\ell_m)$ is the shortest sequence for which  $\ell_1=i$, $\ell_m=j$, and airway $\ell_k$ and airway  $\ell_{k+1}$ are either siblings or one is the parent of the other.        
The equality in \eqref{eq:sumxixj} follows by using that $\uv x_{p}+\uv x_{q}=2\cos(\theta)\uv x_r$ if airway $p$ and airway $q$ are siblings and children of airway $r$.   
By applying the same reasoning to the other terms in the expression, it is seen that all the terms in the expression can be computed by applying the substitution that assigns $\mat C_{ij}$ to $\uv x_i\cdot\uv x_j$. This and the fact that $\mat Q\ex^{-\iu\mat\Theta}=\mat P^\ast$ gives
\begin{align}
\fra{6t}&\mean{{(\vec o_{l_t}+x_t\uv x_{l_t}-\vec o_{l_0}-x_0\uv x_{l_0})^2}}\nonumber\\
&\phantom{}=\frac{\chi L^2}{t} +\fra{t}\sum_{n=1}^\infty a_nL^2\exp(-\alpha_n^2 D_a t/L^2),
\label{eq:firstpart}
\end{align}
where  
\begin{align*}
\chi&=\fra{3E}\sum_i(\mat O^\top\mat C\mat O + \mat C\mat O+ \tfra3\mat C)_{ii}\\
&\,-\fra{3E^2}\sum_{i,i'}(\mat O^\top\mat C\mat O+\mat C\mat O+\tfra4\mat C)_{{i'}i},\\
a_n&=-\fra{3E\alpha_n^2}(\mat S^\top\mat C\mat S +2\mat S^\top\mat C\mat T + \mat T^\top\mat C\mat T)_{r_n r_n}\\
&\,-\!\frac{2}{3E\alpha_n^3}(\mat S^\top\mat C\mat U\!+\!\mat T^\top\mat C\mat U)_{r_nr_n}\!-\!\fra{3E\alpha_n^4}\!(\mat U^\top\mat C\mat U)_{r_nr_n},
\end{align*}
and       
\begin{align*}
\mat S&=2\mat O\Re(\iu\mat P^\ast-\iu\mat Q),\\ 
\mat T&=2\Re(\iu\mat P^\ast),\\
\mat U&=2\Re(\mat P^\ast-\mat Q).
\end{align*}
The $\mat O$ and $\mat C$ matrices can be computed by using that     
\begin{align*}
\mat O&=\sum_{n=1}^{\nu}(\mat\Psi\mat\Phi^\top)^n,\nonumber\\
\mat F&=\left[\begin{array}{cc}
\cos(\theta)\,\mat\Phi\mat\Psi^\top &  \cos(2\theta)(\mat\Phi\mat\Phi^\top -\mat I)\\
\cos(2\theta)(\mat\Psi\mat\Psi^\top-\mat I) & \cos(\theta)\,\mat\Psi\mat\Phi^\top
\end{array}\right],\nonumber\\
\mat C&=\mat I+\sum_{n=1}^{2\nu}(\mat F^n)_{\mat{11}}+(\mat F^n)_{\mat{12}}+(\mat F^n)_{\mat{21}}+(\mat F^n)_{\mat{22}},
\end{align*} 
where the first, second, third, and fourth matrices to the right of the summation sign in the last equation above are the upper-left, upper-right, lower-left, and lower-right $E\times E$ blocks of $\mat F^n$,  respectively. 
The equality in the last equation above follows by using that the geometric graph is acyclic and  $\mat F_{ij}= c_{ij}\delta_{t_i h_j}(1-\delta_{h_i t_j})$, where $t_{i}$ and $h_i$ are equal to $\phi_i$ and $\psi_i$, respectively, if $1\le i\le E$ and equal to $\psi_{i-E}$ and $\phi_{i-E}$, respectively, if $E<i\le 2E$ and $c_{ij}=\uv x_{i-\lfloor(i-1)/E\rfloor E}\cdot\uv x_{j-\lfloor (j-1)/E\rfloor E}$.  

Using, for the second part of $\mathcal D(t)$, the transverse part of the cylindrical-coordinate  continuum-limit propagator for particle diffusion in a nonabsorbing hollow cylinder with radius $R$, i.e.,   
\begin{align*} 
\bar K&(r',\varphi',r,\varphi,t)\\
&=\fra{\pi R^2}+\!\sum_{m=1}^\infty\sum_{n=1}^\infty\frac2{\pi R^2(1-m^2/\beta_{mn}^2)J_m(\beta_{mn})^2}\\
&{\;\:\:\,\,\,}\times\ex^{-\beta_{mn}^2D_bt/R^2}J_m(\beta_{mn}r'/R)J_m(\beta_{mn}r/R)\\
&{\;\:\:\,\,\,}\times\cos(m\varphi'-m\varphi),
\end{align*} 
where $J_m(x)$ is the $m$th-order Bessel function of the first kind and $\beta_{mn}$ is the $n$th root of the derivative of $J_m(x)$ (see, e.g., Refs.~\cite{Neuman1974,Yablonskiy2002}), we get     
\begin{align}
\label{eq:secondpart} 
\fra{6t}&\mean{(y_t\uv y_{l_t}+z_t\uv z_{l_t}-y_0\uv y_{l_0}-z_0\uv z_{l_0})^2}\nonumber\\
&=\fra{6t}\mean{(y_t\uv y_{l_t}+z_t\uv z_{l_t})^2}+\fra{6t}\mean{(y_0\uv y_{l_0}+z_0\uv z_{l_0})^2}\nonumber\\ 
&{\;\:\:\,\,\,}-\frac{2}{6t}\sum_{i}\int_0^R\!\!\int_0^{2\pi}\!\!\int_0^R\!\!\int_0^{2\pi}\frac{1}{E\pi R^2}\bar K(r',\varphi',r,\varphi,t)\nonumber\\
&{\;\:\:\,\,\,}\times(\cos\varphi'\uv y_i+\sin\varphi'\uv z_i)\cdot(\cos\varphi\uv y_i+\sin\varphi\uv z_i)\nonumber\\
&{\;\:\:\,\,\,}\times r'^2r^2\du\varphi\du r\du\varphi'\du r'\nonumber\\
&=\frac{R^2}{6t}+\fra{t}\sum_{n=1}^\infty b_nR^2\exp({-\beta_{1n}^2 D_bt/R^2}),
\end{align} 
where {$b_n=4/(3\beta_{1n}^2-3\beta_{1n}^4)$} 
and we used that  $\int_0^{2\pi}\int_0^{2\pi}\cos(m(\varphi'-\varphi))\sin(\varphi')\sin(\varphi)\du\varphi\du\varphi' =\pi^2\delta_{\abs{m}1}$.  

By inserting the sum of \eqref{eq:firstpart} and \eqref{eq:secondpart} in the integral expression for $\mathfrak D$ we get for $G(t)$ consisting of two rectangular pulses    
\begin{gather*}
\mathfrak D=
\sum_{n=1}^\infty
a_n L^2f(-\alpha_n^2 D_a/L^2)+\sum_{n=1}^\infty b_nR^2 f(-\beta_{1n}^2D_b/R^2),
\end{gather*}
where
 \begin{align*}
f(\kappa)=\frac{2\kappa\delta+2-2\ex^{\kappa\Delta}-2\ex^{\kappa\delta}+\ex^{\kappa(\Delta-\delta)}+\ex^{\kappa(\Delta+\delta)}}{\kappa^2\delta^2(\Delta-\delta/3)}.
\end{align*}

\section*{Appendix B}\label{app:B}
\renewcommand{\theequation}{B1} %
In this Appendix, we consider the behavior of $\mathfrak D$ for values of $\Delta$ such that $D_a\Delta/L^2\sim 1$ and $D_b\Delta/R^2\sim 1$. 
From the general analytic expression for the mean squared displacement for nonfree particle diffusion derived in Ref.~\cite{Buhl2018asymp} we get 
\begin{align}\label{eq:asymptotic}
\fra{6t}&\mean{{(\vec o_{l_t}+x_t\uv x_{l_t}-\vec o_{l_0}-x_0\uv x_{l_0})^2}},\nonumber\\
&\simeq\frac{D_a}3\Big(1-\frac{4}{3\sqrt\pi}\Upsilon\sqrt{D_at}\Big),
\quad D_a t/L^2\lesssim 1,  
\end{align}
where 
\begin{gather*}
\Upsilon=\fra{EL}\sum_{\ell=1}^{V}\fra{d_\ell}\Big(\sum_{i\in N_\ell}s_{i \ell}\uv x_{i}\Big)^2,
\end{gather*}
in which $N_\ell=\{i:\phi_i=\ell\vee\psi_i=\ell\}$, $d_\ell$ is the degree of vertex $\ell$ (i.e., the number of elements in $N_\ell$), and $s_{i\ell }=\delta_{\phi_i \ell}-\delta_{\psi_i \ell}$. 
By inserting the sum of \eqref{eq:asymptotic} and \eqref{eq:secondpart} in the integral expression for $\mathfrak D$ we get for $G(t)$ consisting of two rectangular pulses and $\delta=\Delta$  
\begin{align*} 
\mathfrak D\simeq &\frac{D_a}3\Big(1-\tfrac{257}{511 L}\frac{32(2\sqrt2-1)}{35\sqrt{\pi}}(D_a\Delta)^{1/2}\Big)
+\tfrac{7}{48}\frac{R^4}{D_b\Delta^2}\\
&-\tfrac{33}{512}\frac{R^6}{D_b^2\Delta^3},\quad\!{D_a\Delta}/{L^2}\sim 1,{D_b\Delta}/{R^2}\sim 1,  
\end{align*}
where it is used that   $\sum_{n=1}^\infty\frac{1}{\beta_{1n}^{4}(\beta_{1n}^2-1)}=\tfrac{7}{192}$ and  $\sum_{n=1}^\infty\frac{1}{\beta_{1n}^{6}(\beta_{1n}^2-1)}=\tfrac{11}{1024}$ see, e.g., ($75$) and ($76$) in Ref.~\cite{Sneddon1960} (in ($76$) the first parenthesis in the denominator should be raised to the third power and not the second power as printed).

%

\vspace{-0.1cm}
\onecolumngrid   

\end{document}